\documentclass[12pt]{article}
\usepackage{amssymb,amsmath,amsthm,latexsym}

\newtheorem{thm}{Theorem}[section]
\newtheorem{prop}[thm]{Proposition}
\newtheorem{lem}[thm]{Lemma}
\newtheorem{cor}[thm]{Corollary}
\newtheorem{exam}{Example}
\newtheorem{defi}[thm]{Definition}
\newtheorem{rem}[thm]{Remark}
\newcommand{\pf}{{\bf Proof. \ }}

\begin{document}

\title{Matrix-Product Complementary dual Codes }
\author{
 Xiusheng Liu\\
 School of Mathematics and Physics, \\
 Hubei Polytechnic University  \\
 Huangshi, Hubei 435003, China, \\
{Email: \tt lxs6682@163.com} \\
Hualu Liu\\
 School of Mathematics and Physics, \\
 Hubei Polytechnic University  \\
 Huangshi, Hubei 435003, China, \\
{Email: \tt hwlulu@aliyun.com} \\}
\maketitle


\begin{abstract} Linear complementary dual codes (LCD) are linear codes satisfying $C\cap C^{\perp}=\{0\}$. Under suitable conditions, matrix-product codes that are complementary dual codes are characterized. We construct LCD codes using quasi-orthogonal matrices. Some asymptotic results are derived.
\end{abstract}


\bf Key Words\rm : LCD codes; Matrix-product codes;  dual codes

\section{Introduction}
Linear complementary dual codes (which is abbreviated to LCD codes) are linear codes that meet their dual trivially. These codes were introduced by Massey in [3] and showed that asymptotically good LCD codes exist, and provide an optimum linear coding solution for the two-user binary adder channel. They are also used in counter measure to passive and active side channel analyses on embedded cryto-systems(See[2]). It was proven by Sendrier [5] that LCD codes meet the Gilbert-Varshamov bound. Necessary and sufficient conditions for cyclic codes  to be LCD codes were obtained(See[6]).

Matrix-product codes  over finite fields were introduced in [1]. Many well-known constructions can be formulated as matrix-product codes, for example, the $(a|a + b)$-construction and the $(a + x|b + x|a + b + x)$-construction, and some quasi-cyclic codes can be rewritten as matrix-product codes(See[1,8]). In [4], the authors  studied  complementary-dual quasi-cyclic  codes. Necessary and sufficient conditions for certain classes of quasi-cyclic codes  to be LCD codes were obtained. Motivated by this work, we will investigate necessary and sufficient conditions for certain classes of matrix-product codes
to be LCD codes.

We finish this introduction with a description of each section in this paper. Section 2 reviews the basics about linear codes, matrix-product codes and LCD codes. In section 3,  we propose new constructions of LCD codes by using matrix-product codes,  and, in section 4, using methods of the section 3, concrete examples are presented to construct good parameters LCD codes of the
section 3.  Finally, a brief summary of this work is described in section 5.
\section{Preliminaries}
In order for the exposition in this paper to be self-contained, we introduce some basic concepts and results about linear codes, LCD codes and matrix-product
codes. For more details, we refer to [1,3,10].

Starting from this section till the end of this paper, we assume that $\mathbb{F} _{q}$ is a finite field with $q=p^{m}$ elements, where $p$ is a prime number and $m \geq 1$ is an integer. For a positive integer $n$, let $\mathbb{F}_{q}^{n}$ denote the vector space of all $n$-tuples over $\mathbb{F} _{q}$. A linear $[n,k]_{q}$ code $C$ over $\mathbb{F} _{q}$ is  a $k$-dimensional subspace of $\mathbb{F}_{q}^{n}$. The Hamming weight $w_H(c)$ of a codeword $c \in C$ is the number of nonzero components of $c$. The Hamming distance of two codewords $c_1,c_2 \in C$  is $d_H(c_1,c_2) = w_H(c_1-c_2)$. The minimum Hamming distance $d(C)$ of $C$ is the minimum Hamming distance between any two distinct codewords of $C$. An $[n,k,d]_{q}$ code is an $[n,k]_{q}$ code with the minimum Hamming distance $d$.

For two vectors $a=(a_1,a_2,\ldots,a_n)$ and $b=(b_1,b_2,\ldots,b_n)$ in $\mathbb{F}_{q}^{n}$,
we define the  Euclidean inner product $<a,b>$ to be $<a,b>=\sum_{i=1}^{n}a_{i}b_{i}$.  For a linear $[n,k]_{q}$ code $C$ over $\mathbb{F} _{q}^{n}$,
we define the  Euclidean dual code as
$$C^{\perp}=\{a\in\mathbb{F} _{q}^{n}\mid<a,b>=0~\mathrm{for}~\mathrm{all}~b\in C\}.$$

To define the Hermitian inner product, we will assume that $m=2k$ and $l=p^k$. Then $\mathbb{F} _{q}=\mathbb{F} _{l^{2}}$.
For two vectors $a=(a_1,a_2,\ldots,a_n)$ and $b=(b_1,b_2,\ldots,b_n)$ in $\mathbb{F}_{l^{2}}^{n}$,
we define the Hermitian inner product $<a,b>_{h}$ to be $<a,b>_{h}=\sum_{i=1}^{n}a_{i}b_{i}^{l}$.  For a linear $[n,k]_{l^{2}}$ code $C$ over $\mathbb{F} _{l^{2}}^{n}$,
we define the Hermitian dual code as
$$C^{\perp_{h}}=\{a\in\mathbb{F} _{l^{2}}^{n}\mid<a,b>_{h}=0~\mathrm{for}~\mathrm{all}~b\in C\}.$$
From the fact that the Hermitian inner product is nondegenerate, it
follows immediately that $(C^{\perp_{h}})^{\perp_{h}}=C$ holds. Moreover, one
has $\mathrm{dim}_{\mathbb{F}_{l^{2}}}C^{\perp_{h}}+\mathrm{dim}_{\mathbb{F}_{l^{2}}}C=n$.

A  linear code $C$ is called Hermitian LCD if $C^{\perp_{h}}\cap C=\{0\} $ .

Given a vector $a=(a_1,a_2,\ldots,a_n)$,  we  define the $l$th power of $a$ as
$$a^{l}=(a_1^{l},a_2^{l},\ldots,a_n^{l}).$$
For a linear code $C$ of length $n$ over $\mathbb{F} _{l^{2}}^{n}$, we define $C^l$ to be the
set $\{a^{l}\mid~\mathrm{for}~\mathrm{all}~a\in C\}$.  Then,  it is easy to see that for a linear code $C$ of length $n$ over $\mathbb{F} _{l^{2}}^{n}$,
the Hermitian dual $C^{\perp_{h}}$ is equal to the Euclidean $(C^{l})^{\perp}$ dual of $C^{l}$. Therefore,
$C$ is a Hermitian LCD code if and only if $(C^{l})^{\perp}\cap C=\{0\}$.

\subsection {The Matrix-Product Codes}
The matrix-product codes $C=[C_1,\ldots,C_s]A$  consists of all matrix product $[c_1,\ldots,c_s]A$ where $c_i\in C_i$ is an $n\times1$ column vector and $A=(a_{ij})_{s\times m}$ is an $s\times m$ matrix over $\mathbb{F}_q$. Here $s \leq m$ and $C_i$ is an $(n,\mid C_i\mid,d_i)_{q}$ code over $\mathbb{F}_q$.  If $C_1,\ldots,C_s$ are linear with generator matrices $G_1,\ldots,G_s$, respectively, then $[C_1,\ldots,C_s]A$ is linear with generator matrix
$$G=\begin{pmatrix}a_{11}G_1&a_{12}G_1&\cdots&a_{1m}G_1\\ a_{21}G_2& a_{22}G_2 & \cdots&a_{2m}G_2 \\ \vdots &\vdots&\cdots& \vdots\\a_{s1}G_s & a_{s2}G_s& \cdots & a_{sm}G_s\end{pmatrix}.$$

A matrix-product code $C=[C_1,\ldots,C_s]A$ over $\mathbb{F}_q$ is said to be a matrix-product LCD (MPLCD) code  if  the matrix-product code $C=[C_1,\ldots,C_s]A$ is an LCD code over $\mathbb{F}_q$.

Let $A=(a_{ij})_{s\times m}$ be a matrix over $\mathbb{F}_q$. For any index $1\leq k \leq s$, we denote by $U_{A}(k)$ the linear code of length $m$ over $\mathbb{F}_q$ generated by the $i$th rows of $A$, for $i=1,2,\ldots,k$.
Let $A_t$ be the matrix consisting of the first $t$ rows of  $A=(a_{ij})_{s\times m}$ . For $1\leq j_1 <j_2<\cdots<j_t\leq m$,
we write $A(j_1,j_2,\ldots,j_t)$ for the $t\times t$ matrix consisting of columns $j_1,j_2,\ldots,j_t$ of $A_t$.
\begin{defi} $[1,\mathrm{Definition}~3.1 ]$ Let $A=(a_{ij})_{s\times l}$ be a matrix over $\mathbb{F}_q$.

$(1)$ If the rows of $A$ are linearly independent, then  we say $A$ is a full-row-rank(FRR) matrix.

$(2)$ We call $A$ non-singular by columns (NSC) if $A(j_1,j_2,\ldots,j_t)$ is non-singular for each $1\leq t \leq s$ and  $1\leq j_1 <j_2<\cdots <j_t\leq m$.
\end{defi}

The codes in this paper are based on matrix-product codes, whose parameters are described in the following result.
\begin{lem}$[7]$ If $C_i$ is a linear code over $\mathbb{F}_q$ with parameters $[n,k_i,d_i]_{q}$  for $1\leq i\leq s$, $A=(a_{ij})_{s\times m}$ is an FRR matrix over $\mathbb{F}_q$ and $C=[C_1,\ldots,C_s]A$, then $C$ is an $[nm,\sum_{i=1}^{s}k_i,d(C)]_{q}$ linear code over $\mathbb{F}_q$ whose minimum distance $d(C)$ satisfying
$$d(C)\geq \mathrm{min}\{d_1d(U_A(1)),d_2d(U_A(2)),\ldots,d_sd(U_A(s)\}.$$
\end{lem}

\begin{lem}$[1,\mathrm{Theorem}~ 3.7, \mathrm{Proposition}~ 6.2]$ If $C_i$ is a linear code over $\mathbb{F}_q$ with parameters $[n,k_i,d_i]_{q}$ for $1\leq i\leq s$, $A$ is an $s\times m$ NSC matrix and $C=[C_1,\ldots,C_s]A$, then

$(1)$ $d(C)\geq d^{*}=\mathrm{min}\{md_1,(m-1)d_2,\ldots,(m-s+1)d_s\}$;

$(2)$ If $A$ is also upper-triangular then $d(C)= d^{*}$.
\end{lem}

\begin{lem} $[1,\mathrm{Theorem}~6.6, \mathrm{Proposition}~ 6.2]$ Let  $A$ be an  $s\times s$ non-singular matrix and $C_1,\ldots,C_s$ are linear codes over $\mathbb{F}_q$. If $C=[C_1,\ldots,C_s]A$, then

$(1)$ $([C_1,\ldots,C_s]A)^{\perp}=[C_1^{\perp},\ldots,C_s^{\perp}](A^{-1})^{T}$;

$(2)$ If $A$ is also $s\times s$ NSC matrix, then $d(C^{\perp })\geq (d^{\perp})^{*}=\mathrm{min}\{sd_s^{\perp},(s-1)d_{s-1}^{\perp},\ldots,d_1^{\perp}\}$ where $d_j^{\perp}=d(C_j^{\perp})$, for $j=1,2,\ldots,s$;

$(3)$ If $A$ is also upper-triangular then $d(C^{\perp})= (d^{\perp})^{*}$.
\end{lem}

\subsection {Cyclic LCD Codes}
We first reviews the basics about Euclidean LCD codes.

Given a ring $\widetilde{R}$, for a nonempty subset $S$ of $\widetilde{R}$, the annihilator of $S$, denoted by $ann($S$)$, is the set
$$ann(S)=\{f\in\widetilde{R}|fg=0,\,\,\mbox{for\, all}\,\, g \in S\}.$$
If, in addition, $S$ is an ideal of $\widetilde{R}$, then $ann(S)$ is also an ideal of $\widetilde{R}$.

For any polynomial $f(x)=\sum_{i=0}^{k}a_{i}x^{i}$ of degree $k$ ($a_{k}\neq0)$ over $\mathbb{F}_{q}$, let $f^{\ast}(x)$ denote the reciprocal polynomial of $f(x)$ given by
\\$$f^{\ast}(x)=x^{k}f(\frac{1}{x})=\sum_{i=0}^{k}a_{k-i}x^{i}.$$
Note that $(f^{\ast})^{\ast}=f$ if and only if the constant term of $f$ is nonzero, if and only if deg($f$)=deg($f^{\ast}$). Furthermore, by definition, it is easy to see that  $(fg)^{\ast}=f^{\ast}g^{\ast}$. We denote $N^{\ast}=\{f^{\ast}(x)|f(x)\in N\subset \mathbb{F}_q[x]\}.$ It is easy to see that if $N$ is an ideal, then $N^{\ast}$ is also an ideal. Hereafter, we will use $ann^{\ast}(C)$ to denote$(ann(C))^{\ast}$.

The following Proposition can be found in [10].

\begin{prop}
 If $C$ is a cyclic  code of length $n$ over $\mathbb{F}_{q}$, then the dual $C^{\perp}$ of $C$ is $ann^{\ast}(C)$.
\end{prop}

Suppose that $f(x)$ is a monic (i.e., leading coefficient $1$) polynomial of degree $k$ with $f(0)=c\neq0.$ Then by monic reciprocal  polynomial of $f(x)$  we mean the  polynomial $\widetilde{f}(x)=c^{-1}f^{\ast}(x)$.

We recall a result about cyclic LCD codes which can be found in [6].

\begin{prop}
If $g_{1}(x)$ is the generator polynomial of a cyclic code $C$ of length $n$ over $\mathbb{F}_{q}$, then $C$ is an LCD code if and only if  $g_{1}(x)$ is self-reciprocal $(i.e.,\widetilde{g}_{1}(x)=g_{1}(x))$ and all the monic irreducible factors of $g_{1}(x)$ have the same multiplicity in $g_{1}(x)$ and in $x^{n}-1$.
\end{prop}

In the following we  investigate the generator polynomials of the repeated-root cyclic LCD codes over $\mathbb{F}_{q}$ where $q=p^m$.

It is well known that each cyclic code over $\mathbb{F}_{q}$ is uniquely determined by its generator polynomial, a monic divisor of $x^{n}-1$ over $\mathbb{F}_{q}$. In order to describe the generator polynomials of  the repeated-root cyclic LCD codes, we need to know the factorization of the polynomial $x^{n}-1$ over $\mathbb{F}_{q}$. Write
$$\qquad\qquad\qquad\qquad\qquad n=p^{t}l,\qquad\qquad\qquad\qquad\qquad\qquad\qquad\qquad\qquad\quad\quad$$
where $t$ is a nonnegative integer depending on $n$ and $gcd(l,p)=1.$ Then
$$x^{n}-1=(x^l-1)^{p^{t}}.$$

For any irreducible polynomial dividing $x^{l}-1$ over $\mathbb{F}_{q}$, its reciprocal polynomial also divides  $x^{l}-1$ over $\mathbb{F}_{q}$ and is also irreducible over $\mathbb{F}_{q}$. Since $gcd(l,p)=1$, the polynomial $x^{l}-1$ factors completely into nonassociated irreducible factors in $\mathbb{F}_{q}[x]$ as
$$x^{l}-1=\delta f_{1}(x)f_{2}(x)\cdots f_{k}(x)h_{1}(x)h_{1}^{\ast}(x)\cdots h_{s}(x)h_{s}^{\ast}(x),$$
where $\delta\neq0$ in $\mathbb{F}_{q}$, $f_{1}(x),f_{2}(x), \cdots, f_{k}(x)$ are irreducible polynomials that are associates to their own reciprocals, and $h_{1}(x),h_{1}^{\ast}(x);\cdots; h_{s}(x),h_{s}^{\ast}(x)$ are pairs of mutually reciprocal irreducible polynomials. Therefore
$$ x^{n}-1= \delta^{p^{t}} (f_{1}(x))^{p^{t}}(f_{2}(x))^{p^{t}}\cdots (f_{k}(x))^{p^{t}}(h_{1}(x))^{p^{t}}(h_{1}^{\ast}(x))^{p^{t}}\cdots (h_{s}(x))^{p^{t}}(h_{s}^{\ast}(x))^{p^{t}}.~~~(2.1)$$
We can describe the generator polynomials of the repeated-root cyclic LCD codes as soon as we know the factorization of $x^{n}-1$ over $\mathbb{F}_{q}$.
\begin{prop}
Let $x^{n}-1$ be factorized as in $(2.1)$. A repeated-root cyclic code $C$ of length $n$ over $\mathbb{F}_{q}$ is an LCD code if and only if its generator polynomial is of the form
$$ (f_{1}(x))^{\alpha_{1}}(f_{2}(x))^{\alpha_{2}}\cdots (f_{k}(x))^{\alpha_{k}}(h_{1}(x))^{\beta_{1}}(h_{1}^{\ast}(x))^{\beta_{1}}\cdots (h_{s}(x))^{\beta_{s}}(h_{s}^{\ast}(x))^{\beta_{s}}.\qquad\qquad(2.2)$$
where $\alpha_{i}\in\{0,p^t\}$ for each $1\leq i \leq k$, and $ \beta_{j}\in\{0,p^t\}$ for each $1\leq j \leq s$ .
\end{prop}
\pf
Let $C$ be a cyclic code of length $n$ over $\mathbb{F}_{q}$, and let $g(x)$ be its generator polynomial. We need to show that $C$ is an LCD code if and only if  $g(x)$ is of the form as in (2.2).

Suppose that
$$ g(x)=\varepsilon(f_{1}(x))^{\alpha_{1}}(f_{2}(x))^{\alpha_{2}}\cdots (f_{k}(x))^{\alpha_{k}}(h_{1}(x))^{\beta_{1}}(h_{1}^{\ast}(x))^{\gamma_{1}}\cdots (h_{s}(x))^{\beta_{s}}(h_{s}^{\ast}(x))^{\gamma_{s}}$$
with leading coefficient 1, where $0 \leq \alpha_{i} \leq p^t$ for each $1\leq i \leq k$, and $0\leq \beta_{j},\gamma_{j}\leq p^t$ for each $1\leq j \leq s$, where $\varepsilon\in\mathbb{F}_{q}$. Then
$$ g^{\ast}(x)=\eta(f_{1}(x))^{\alpha_{1}}(f_{2}(x))^{\alpha_{2}}\cdots (f_{k}(x))^{\alpha_{k}}(h_{1}(x))^{\gamma_{1}}(h_{1}^{\ast}(x))^{\beta_{1}}\cdots (h_{s}(x))^{\gamma_{s}}(h_{s}^{\ast}(x))^{\beta_{s}},$$
where $\eta\in\mathbb{F}_{q}$.
Therefore,
$$\widetilde{g}(x)=\frac{1}{g(0)}g^{\ast}(x)=\varepsilon(f_{1}(x))^{\alpha_{1}}(f_{2}(x))^{\alpha_{2}}\cdots (f_{k}(x))^{\alpha_{k}}(h_{1}(x))^{\gamma_{1}}(h_{1}^{\ast}(x))^{\beta_{1}}\cdots (h_{s}(x))^{\gamma_{s}}(h_{s}^{\ast}(x))^{\beta_{s}}.$$

By Proposition 2.6, $C$  is an LCD code if and only if $g(x)=\widetilde{g}(x)$ and all the monic irreducible factors of $g(x)$ have the same multiplicity in  $g(x)$  and in $x^{n}-1$, i.e., $\beta_{j}=\gamma_{j}$ for each  $1\leq j \leq s$, $\alpha_{i}\in\{0,p^t\}$ for each $1\leq i \leq k$, and $ \beta_{j}\in\{0,p^t\}$ for each $1\leq j \leq s$.

Therefore, $C$ is an LCD code if and only if its generator polynomial $g(x)$ is of the form as in (3.2).
\qed

Next, we reviews the basics about Hermitian LCD codes which can be found in [8].

Let $f(x) = a_{0} +a_{ 1}x+\ldots+a_{k}x^{k}\in\mathbb{F}_{l^{2}}[x]$ with a nonzero constant coefficient $a_0$, let us denote the monic
polynomial by

$$f^{\perp}(x)= a_{0}^{-l}(a_{k}^{l} +a_{k-1}^{l}x+\ldots+a_{0}^{l}x^{k}).$$
Then $f^{\perp}(x)$ is called  conjugate-reciprocal  polynomial of $f(x)$.

\begin{lem} $[9,\mathrm{Lemma}~7]$ Assume that $C= \langle g(x)\rangle$ is a cyclic code of length $n$ over $\mathbb{F}_{l^{2}}$ with generator
polynomial $g(x)$. Define $h(x)=\frac{x^n-1}{g(x)}$. Then we have that $C^{\perp_{h}} = \langle h^{\perp}(x)\rangle$.
\end{lem}

Similar to the discussions in [3], we have the following proposition.

\begin{prop}
If $g(x)$ is the generator polynomial of a cyclic code $C$ of length $n$ over $\mathbb{F}_{l^{2}}$, then $C$ is an LCD cyclic code with respect to Hermitian inner product if and only if  g(x) is conjugate-self-reciprocal $(\mathrm{i.e.,}g(x)=g^{\perp}(x))$  and all the monic irreducible factors of $g(x)$ have the same multiplicity in $g(x)$ and in $x^{n}-1$.
\end{prop}
\section{Matrix-Product LCD Codes}

Let $A =(a_{ij} )$ be an $s\times m$ matrix with entries in $\mathbb{F}_{q}$.  $A$ is said to be a quasi-orthogonal matrix if $AA^{T}$  is a diagonal square matrix where all the diagonal entries are non-zeros of $\mathbb{F}_{q}$. Then we have the following result.

\begin{thm}
Let  $C_1,C_2,\ldots,C_s$ be  linear codes over $\mathbb{F}_{q}$, and  let  $A$ is  an  $s\times s$  quasi-orthogonal  matrix over $\mathbb{F}_{q}$.  Then C=$[C_1,C_2,\ldots,C_s]A$ is an  MPLCD code if and only if  $C_1,C_2,\ldots,C_s$ are all LCD codes. In particular, if $(A^{-1})^{T}=aA$, where $a\in\mathbb{F}_{q}^{*}$, then $C=[C_1,C_2,\ldots,C_s]A$ is an MPLCD code if and only if  $C_1,C_2,\ldots,C_s$ are all LCD codes.
\end{thm}
\pf Assume that $AA^{T}=\begin{pmatrix}r_{1}&0&\cdots&0\\ 0& r_{2} & \cdots&0\\ \vdots &\vdots&\cdots& \vdots\\0 & 0& \cdots & r_{s}\end{pmatrix}$, with $r_i$ being non-zeros of $\mathbb{F}_{q}$. Then $$(A^{-1})^{T}=\begin{pmatrix}r_{1}^{-1}&0&\cdots&0\\ 0& r_{2}^{-1} & \cdots&0\\ \vdots &\vdots&\cdots& \vdots\\0 & 0& \cdots & r_{s}^{-1}\end{pmatrix}A.$$ Therefore, by Lemma 2.4, the dual code
$$~~~~~~~~~~~~~~~~~C^{\perp}=([C_1,\ldots,C_s]A)^{\perp}~~~~~~~~~~~~~~~~~~~~~~~~~~~~~~~~~~~~~~~~~~$$
$$~~~~~~~~~~~~~~~~~=([C_1^{\perp},\ldots,C_s^{\perp}](A^{-1})^{T}~~~~~~~~~~~~~~~~~~~~~~~~~~~~~~~~~$$
$$~~~~~~~~~~~~~~~~~~~~~=[C_1^{\perp},\ldots,C_s^{\perp}] \begin{pmatrix}r_{1}^{-1}&0&\cdots&0\\ 0& r_{2}^{-1} & \cdots&0\\ \vdots &\vdots&\cdots& \vdots\\0 & 0& \cdots & r_{s}^{-1}\end{pmatrix}A.~~~~~~~~~~~~~~~$$
Since $C_j$ is a linear code, so also is $C_j^{\perp}$. This implies that $r_j^{-1}C_j^{\perp}=C_j^{\perp}$, for $j=1,2,\ldots,s$.
Hence,
$$~~~~~~~~~~~C^{\perp}=[C_1^{\perp},\ldots,C_s^{\perp}] \begin{pmatrix}r_{1}^{-1}&0&\cdots&0\\ 0& r_{2}^{-1} & \cdots&0\\ \vdots &\vdots&\cdots& \vdots\\0 & 0& \cdots & r_{s}^{-1}\end{pmatrix}A=[C_1^{\perp},\ldots,C_s^{\perp}]A.~~~~~~~~~~~~~~(3.1)$$

$\Rightarrow:$ If $C$ is not an MPLCD code over $\mathbb{F}_{q}$, then there exists a nonzero codeword $\alpha$ in $C$ such that $\alpha=(c_1,\ldots,c_s)A\in C\cap C^{\perp}$, where $c_1\in C_1,\ldots,c_s\in C_s$. Obviously, there exists some positive integer $j$ such that $c_j\neq 0$ where $1\leq j\leq s$.

As  $\alpha\in  C^{\perp}$, by above equality (3.1), there exist $c_1^{\perp}\in C_1^{\perp},\ldots,c_j^{\perp}\in C_j^{\perp},\ldots,c_s^{\perp}\in C_s^{\perp}$ such that
$$\alpha=(c_1^{\perp},\ldots, c_j^{\perp},\ldots,c_s^{\perp})A.$$
Therefore
$$(c_1,\ldots,c_j,\ldots,c_s)A=(c_1^{\perp},\ldots,c_j^{\perp},\ldots,c_s^{\perp})A.$$
Since A is an  $s\times s$ non-singular matrix, we obtain
$$(c_1,\ldots,c_j,\ldots,c_s)=(c_1^{\perp},\ldots,c_j^{\perp},\ldots,c_s^{\perp}),$$
which implies that $0\neq c_j=c_j^{\perp}\in C_{j}\cap C_{j}^{\perp}$. This is a contradiction.

$\Longleftarrow:$ Otherwise, we assume that $C_j$ is not an LCD code for some $1\leq j \leq s$. Then there exits a nonzero codeword $c_j$ in $C_j$ such that $ c_j\in C_j^{\perp}$.

Taking $\beta=(0,\ldots,0,c_j,0,\ldots,0)A$, then $\beta\neq0$ since A is an $s\times s$ non-singular matrix. By definition of matrix-product code and above equality (3.1), we have $\beta\in C\cap C^{\perp}$. This is a contradiction.
\qed

\begin{thm}
 Assume that $A$ is  an  $s\times s$ non-singular lower-triangular matrix over $\mathbb{F}_{q}$. If $C_1\supset C_2\supset\cdots\supset C_s$  are  linear codes over $\mathbb{F}_{q}$, then $C=[C_1,C_2,\ldots,C_s]A$ is  an MPLCD code if and only if  $C_1,C_2,\ldots,C_s$ are all LCD codes.
\end{thm}
\pf Suppose that $A=\begin{pmatrix}r_{11}&0&0&\cdots&0\\ r_{21}& r_{22}&0& \cdots&0\\ \vdots &\vdots&\vdots&\cdots& \vdots\\r_{s1} & r_{s2}&r_{s3}& \cdots & r_{ss}\end{pmatrix}$. Then, for $i=1,\ldots,s$, $r_{ii}$ is non-zeros of $\mathbb{F}_{q}$ since $A$ is  an  $s\times s$ non-singular lower-triangular matrix. Therefore
$$(A^{-1})^{T}=\begin{pmatrix}r_{11}^{-1}&r_{12}^{'}&\cdots&r_{1s}^{'}\\ 0& r_{22}^{-1} & \cdots&r_{2s}^{'}\\ \vdots &\vdots&\cdots& \vdots\\0 & 0& \cdots & r_{ss}^{-1}\end{pmatrix}.$$
where $r_{ij}'\in \mathbb{F}_{q},  i<j, i=1,\ldots,s-1, j=2,\ldots,s$.
Hence, by Lemma 2.4, the dual code
$$C^{\perp}=([C_1,\ldots,C_s]A)^{\perp}=[C_1^{\perp},\ldots,C_s^{\perp}](A^{-1})^{T}~~~~~~~~~~~~~~~~~~~~~~~~~~~~~~~~~~~~~~~~~~~~~~~$$
$$~~~~~~~~~~~~~~~~~~~~~~=[C_1^{\perp},\ldots,C_s^{\perp}] \begin{pmatrix}r_{11}^{-1}&r_{12}^{'}&\cdots&r_{1s}^{'}\\ 0& r_{22}^{-1} & \cdots&r_{2s}^{'}\\ \vdots &\vdots&\cdots& \vdots\\0 & 0& \cdots & r_{ss}^{-1}\end{pmatrix}~~~~~~~~~~~~~~~~~~~~~~~$$
$$~~~~~~~~~~~~~~~~~~~~~~~~~~~~=[r_{11}^{-1}C_1^{\perp},r_{12}'C_1^{\perp}+r_{22}^{-1}C_2^{\perp},\ldots,r_{1s}'C_1^{\perp}+r_{2s}'C_2^{\perp}+\ldots+r_{ss}^{-1}C_s^{\perp}].$$
Since  $C_1\supset C_2\supset\cdots\supset C_s$  are  linear codes, so $C_1^{\perp}\subset C_2^{\perp}\subset\cdots\subset C_s^{\perp}$ also are linear codes. Then we obtain
$$r_{11}^{-1}C_1^{\perp}=C_1^{\perp},~~r_{12}'C_1^{\perp}+r_{22}^{-1}C_2^{\perp}=C_2^{\perp},~~\ldots,~~r_{1s}'C_1^{\perp}+r_{2s}'C_2^{\perp}+\ldots+r_{ss}^{-1}C_s^{\perp}=C_s^{\perp}.$$
Hence,
$$~~~~~~~~~~~C^{\perp}=[C_1^{\perp},\ldots,C_s^{\perp}] \begin{pmatrix}r_{11}^{-1}&r_{12}^{'}&\cdots&r_{1s}^{'}\\ 0& r_{22}^{-1} & \cdots&r_{2s}^{'}\\ \vdots &\vdots&\cdots& \vdots\\0 & 0& \cdots & r_{ss}^{-1}\end{pmatrix}=[C_1^{\perp},\ldots,C_s^{\perp}].~~~~~~~~~~~~~~(3.2)$$

$\Rightarrow:$ If $C$ is not an MPLCD code over $\mathbb{F}_{q}$, then there exists a nonzero codeword $\alpha=(c_1,\ldots,c_s)A$ in $C$ such that $\alpha\in C^{\perp}$, where $c_1\in C_1,\ldots,c_s\in C_s$. Obviously, there exists some positive integer $j$ such that $c_j\neq 0$ where $1\leq j\leq s$.

As  $\alpha\in  C^{\perp}$, by above equality (3.2), there exist $c_1^{\perp}\in C_1^{\perp},\ldots,c_j^{\perp}\in C_j^{\perp},\ldots,c_s^{\perp}\in C_s^{\perp}$ such that
$$\alpha=(c_1^{\perp},\ldots,c_j^{\perp},\ldots,c_s^{\perp}).$$
Therefore
$$(c_1,\ldots,c_j,\ldots,c_s)A=(c_1^{\perp},\ldots,c_j^{\perp},\ldots,c_s^{\perp}),$$
or
$$(r_{11}c_1+\cdots+r_{s1}c_s,\ldots,r_{jj}c_j+\cdots+r_{sj}c_s,\ldots,r_{ss}c_s)=(c_1^{\perp},\ldots,c_j^{\perp},\ldots,c_s^{\perp}).$$
Since  $C_1\supset C_2\supset\cdots\supset C_s$  are  linear codes,  we obtain
$$r_{11}c_1+\cdots+r_{s1}c_s=c_1'\in C_1,~~\ldots,~~r_{jj}c_j+\cdots+r_{sj}c_s=c_j'\in C_j,~~\ldots,~~r_{ss}c_s=c_s'\in C_s.$$
This implies that $0\neq c_j'=c_j^{\perp}\in C_j\cap C_j^{\perp}$. This is a contradiction.

$\Longleftarrow:$ Otherwise, we assume that $C_j$ is not an LCD code for some $1\leq j \leq s$. Then there exits a nonzero codeword $c_j$ in $C_j$ such that $ c_j\in  C_j^{\perp}$.

Taking $\beta=(0,\ldots,0,c_j,0,\ldots,0)$, then $\beta\neq0$ and $\beta\in C^{\perp}$ by $(\mathrm{3.2})$. Since A is an $s\times s$ non-singular matrix, linear equations over $\mathbb{F}_{q}$
 $$\left\{\begin{aligned}
         &x_{1}r_{11}+x_{2}r_{21}+x_{3}r_{31}+\cdots+x_{j}r_{j1}=0,\\
         &~~~~~~~~~~x_{2}r_{22}+x_{3}r_{32}+\cdots+x_{j}r_{j2}=0,\\
         &\ldots\ldots\ldots\ldots\ldots\ldots\ldots\ldots\ldots\ldots\ldots\ldots\\
         &~~~~~~~~~~~~~~~~x_{j-1}r_{j-1,j-1}+x_{j}r_{j,j-1}=0,\\
         &~~~~~~~~~~~~~~~~~~~~~~~~~~~~~~~~~~~~~~x_{j}r_{jj}=1.
         \end{aligned}\right. $$
have unique solution $x_{1}=v_1,x_2=v_2,\ldots,x_j=v_j$.

It is easy to check that $\beta=(v_1c_j,v_2c_j,\ldots,v_{j-1}c_j,v_jc_j,0,\ldots,0)A$. Hence, $\beta\in C$, i.e., $\beta\in C\cap C^{\perp}$, which is a contradiction.
\qed

Let $A = (a_{ij} )$ be an $s\times s$ matrix with entries in $\mathbb{F}_{l^{2}}$ , we define $A^{(l)} =(a^{l}_{ij})$.
Similarly, We can prove the following two theorems.
\begin{thm}
Let  $C_j\subset\mathbb{F}_{l^{2}}^{n}$ be a linear code  over $\mathbb{F}_{l^{2}}$ for $1\leq i\leq s$, and let $C=[C_1,C_2,\ldots,C_s]A$ where $A$ is an $s\times s$ matrix. If $A^{(l)}A^{T}$  is a diagonal square matrix where all the diagonal entries are non-zeros of $\mathbb{F}_{l^{2}}$. Then $C=[C_1,C_2,\ldots,C_s]A$ is a Hermitian LCD code if and only if $C_1,C_2,\ldots,C_s$ are all Hermitian LCD codes. In particular, if $[(A^{(l)})^{-1}]^{T}=aA$, where $a\in\mathbb{F}_{l^{2}}^{*}$, then $[C_1,C_2,\ldots,C_s]A$ is a Hermitian LCD code if and only if $C_1,C_2,\ldots,C_s$ are all Hermitian LCD codes.
\end{thm}
\begin{thm}
 Assume that $A$ is  an  $s\times s$ non-singular lower-triangular matrix over $\mathbb{F}_{l^{2}}$. If $C_1\supset C_2\supset\cdots\supset C_s$  are  linear codes over $\mathbb{F}_{l^{2}}$, then $C=[C_1,C_2,\ldots,C_s]A$ is a Hermitian LCD code if and only if  $C_1,C_2,\ldots,C_s$ are all Hermitian LCD codes.
\end{thm}

In the following,  we will prove the existence of an asymptotically good sequence of MPLCD codes by Theorem 3.1.
\begin{thm}
MPLCD codes over $\mathbb{F}_{q}$ are asymptotically good.
\end{thm}
\pf It is well known that LCD codes over $\mathbb{F}_{q}$ are asymptotically good (See [3]). let $C_1,\ldots,C_j,\ldots,$ be an asymptotically good sequence of LCD codes over $\mathbb{F}_{q}$, and let $C_j$ be an $[n_{j},k_{j},d_{j}]$ code over $\mathbb{F}_{q}$. We have two case.

Case 1. $q=p^{m}$ where $p$ is an odd prime. In the case,  For each $j\geq 1$, we define the matrix-product code $M_j$ as
$$~~~~~~~~~~~~~~~~~~~~~~~~~~M_j:=[C_j,C_j]\begin{pmatrix}1&1\\ 1&p-1 \end{pmatrix}.~~~~~~~~~~~~~~~~~~~~~~~~~~~~~~~~~~~~~~~~(3.3)$$
Since $\begin{pmatrix}\begin{pmatrix}1&1\\ 1&p-1 \end{pmatrix}^{-1}\end{pmatrix}^{T}=\frac{1}{2}\begin{pmatrix}1&1\\ 1&p-1 \end{pmatrix}$,  $M_j$ is an MPLCD code over $\mathbb{F}_{q}$ by Theorem 3.1. According to Lemma 2.2,  $M_j$ is an $[2n_{j},2k_{j},\geq d_{j}]_{q}$ MPLCD code over $\mathbb{F}_{q}$ respectively.

For the sequence of MPLCD codes  $M_1,\ldots,M_j,\ldots$, the asymptotic rate is
$$R_1=\lim_{j\rightarrow\infty}\frac{2k_{j}}{2n_{j}}=\lim_{j\rightarrow\infty}\frac{k_{j}}{n_{j}},$$
and this quantity is positive since $C_1,\ldots,C_j,\ldots,$ is asymptotically good. For the asymptotic distance, we have

$$\delta_1=\lim_{j\rightarrow\infty}\frac{d(M_j)}{2n_{j}}\geq \lim_{j\rightarrow\infty}\frac{d_{j}}{2n_{j}}.$$
Note again that $\delta_1$ is positive since $C_1,\ldots,C_j,\ldots,$ is asymptotically good. Proving the theorem in $p$ odd prime.

Case 2.  $q=2^{m}$. In the case,  For each $j\geq 1$, we define the matrix-product code $M_j$ as
$$~~~~~~~~~~~~~~~~~~~~~~~~~~M_j:=[C_j,C_j]\begin{pmatrix}1&1\\ 0&1 \end{pmatrix}.~~~~~~~~~~~~~~~~~~~~~~~~~~~~~~~~~~~~~~~~(3.4)$$
We first prove that if $C_j$ is an LCD code over $\mathbb{F}_{2^{m}}$ then $M_j$ is an MPLCD code over $\mathbb{F}_{2^{m}}$.

In fact, $\begin{pmatrix}\begin{pmatrix}1&1\\ 0&1 \end{pmatrix}^{-1}\end{pmatrix}^{T}=\begin{pmatrix}1&0\\ 1&1 \end{pmatrix}$. Hence, we have
$$M_j^{\perp}=[C_j^{\perp},C_j^{\perp}]\begin{pmatrix}1&0\\ 1&1 \end{pmatrix}.$$

Suppose that $\eta \in M_j\cap M_j^{\perp}$. Then there exist $c_j,c_j^{'}\in C_{j}$ and $c_j^{\perp},(c_j^{'})^{\perp}\in C_{j}^{\perp}$ such that $\eta=(c_j,c_j+c_j^{'})=(c_j^{\perp}+(c_j^{'})^{\perp},(c_j^{'})^{\perp})$ by the definition of matrix-product code. This implies that $c_j=c_j^{\perp}+(c_j^{'})^{\perp}$ and $c_j+c_j^{'}=(c_j^{'})^{\perp}$. By $c_j=c_j^{\perp}+(c_j^{'})^{\perp}$ and $C_j^{\perp}$ linearly, we have $c_j\in C_j\cap C_j^{\perp}=\{0\}$, i.e., $c_j=0$. On other hand, by $c_j=c_j^{\perp}+(c_j^{'})^{\perp}$ and $c_j+c_j^{'}=(c_j^{'})^{\perp}$, we obtain $c_j^{'}=c_j^{\perp}\in C_j\cap C_j^{\perp}=\{0\}$, i.e., $c_j^{'}=0$, which proves $\eta=0$. This means that $M_j$ is an MPLCD code over $\mathbb{F}_{2^{m}}$.

According to Lemma 2.3,  $M_j$ is an $[2n_{j},2k_{j}, d_{j}]_{2^{m}}$ MPLCD code over $\mathbb{F}_{2^{m}}$ respectively.

For the sequence of MPLCD codes  $M_1,\ldots,M_j,\ldots$, the asymptotic rate is
$$R_2=\lim_{j\rightarrow\infty}\frac{2k_{j}}{2n_{j}}=\lim_{j\rightarrow\infty}\frac{k_{j}}{n_{j}},$$
and this quantity is positive since $C_1,\ldots,C_j,\ldots,$ is asymptotically good. For the asymptotic distance, we have

$$\delta_2=\lim_{j\rightarrow\infty}\frac{d(M_j)}{2n_{j}}=\lim_{j\rightarrow\infty}\frac{d_{j}}{2n_{j}}.$$
Note again that $\delta_2$ is positive since $C_1,\ldots,C_j,\ldots,$ is asymptotically good. This proves the expected result.
\qed

Similarly, We can prove the following theorem by Theorem 3.3.

\begin{thm}
Matrix-product Hermitian  LCD codes over $\mathbb{F}_{l^{2}}$ are asymptotically good.
\end{thm}

\section{Examples}
In this section, we first construct a class of matrix $A=(a_{ij})_{s\times s}$ whose satisfy  $[(A^{(l)})^{-1}]^{T}=sA$ (or $(A^{-1})^{T}=sA$) over $\mathbb{F}_{l^{2}}$ (or $\mathbb{F}_{q}$)  , then we will provide examples of LCD codes by Theorem 3.1 3.2, 3.3, and 3.4.

 For $\mathbb{F}_{l^{2}}$ with $l$ an odd prime power, let $G=\underbrace{G_{2}\times G_2\times\cdots\times G_2}_{r}$ be abelian group where $G_2=<a:a^2=1>$. Set $s=2r$, and we assume that $s\mid(l^{2}-1)$. Let $\widehat{G}$ be the set of characters for $G$ with respect to $\mathbb{F}_{l^{2}}$. Then $\widehat{G}=\{\chi_0,\chi_1,\ldots,\chi_{s-1}\}$ where $\chi_0,\chi_1,\ldots,\chi_{s-1}$ are the irreducible characters of $G$. For any $g\in G$, it is easy to see that  $\chi_{j}(g)^{2}=1$ for $j=0,1,\ldots,s-1$. Therefore, the character table of $G$ is
$$ A=\begin{pmatrix}\chi_{0}(g_0)&\chi_{1}(g_0)&\cdots&\chi_{s-1}(g_0)\\\chi_{0}(g_1)&\chi_{1}(g_1)&\cdots&\chi_{s-1}(g_1)\\ \vdots &\vdots&\cdots& \vdots\\\chi_{0}(g_{s-1})&\chi_{1}(g_{s-1})&\cdots&\chi_{s-1}(g_{s-1})\end{pmatrix}.$$
where $g_0,g_1,\ldots,g_{s-1}\in G$.  Suppose that $l=2t+1$ for some positive integer $t$ since $l$ is an odd number. Then
$$ A^{(l)}=\begin{pmatrix}(\chi_{0}(g_0))^{l}&(\chi_{1}(g_0))^{l}&\cdots&(\chi_{s-1}(g_0))^{l}\\(\chi_{0}(g_1))^{l}&(\chi_{1}(g_1))^{l}&\cdots&(\chi_{s-1}(g_1))^{l}\\ \vdots &\vdots&\cdots& \vdots\\(\chi_{0}(g_{s-1}))^{l}&(\chi_{1}(g_{s-1}))^{l}&\cdots&(\chi_{s-1}(g_{s-1}))^{l}\end{pmatrix}.~~~~~~~~~~~$$
$$~~~~~~~=\begin{pmatrix}(\chi_{0}(g_0))^{2t+1}&(\chi_{1}(g_0))^{2t+1}&\cdots&(\chi_{s-1}(g_0))^{2t+1}\\(\chi_{0}(g_1))^{2t+1}&(\chi_{1}(g_1))^{2t+1}&\cdots&(\chi_{s-1}(g_1))^{2t+1}\\ \vdots &\vdots&\cdots& \vdots\\(\chi_{0}(g_{s-1}))^{2t+1}&(\chi_{1}(g_{s-1}))^{2t+1}&\cdots&(\chi_{s-1}(g_{s-1}))^{2t+1}\end{pmatrix}.$$
$$~~~~~~~~~~~=\begin{pmatrix}\chi_{0}(g_0)&\chi_{1}(g_0)&\cdots&\chi_{s-1}(g_0)\\\chi_{0}(g_1)&\chi_{1}(g_1)&\cdots&\chi_{s-1}(g_1)\\ \vdots &\vdots&\cdots& \vdots\\\chi_{0}(g_{s-1})&\chi_{1}(g_{s-1})&\cdots&\chi_{s-1}(g_{s-1})\end{pmatrix}=A.~~~~~~~~~~~~~~~~~~~~~~~$$
It is well-known that $A$ is invertible, and we have
$$ (A^{(l)})^{-1}=A^{-1}=\frac{1}{s}\begin{pmatrix}\chi_{0}(g_0)^{-1}&\chi_{0}(g_1)^{-1}&\cdots&\chi_{0}(g_{s-1})^{-1}\\\chi_{1}(g_0)^{-1}&\chi_{1}(g_1)^{-1}&\cdots&\chi_{1}(g_{s-1})^{-1}\\ \vdots &\vdots&\cdots& \vdots\\\chi_{s-1}(g_{0})^{-1}&\chi_{s-1}(g_{1})^{-1}&\cdots&\chi_{s-1}(g_{s-1})^{-1}\end{pmatrix}.$$
For any $g\in G$, $\chi_{j}(g)^{2}=1$, i.e., $\chi_{j}(g)^{-1}=\chi_{j}(g)$ for $j=0,1,\ldots,s-1$. This implies that $[(A^{(l)})^{-1}]^{T}=\frac{1}{s}A$.

\begin{rem} For $\mathbb{F}_{q}$, let $G=\underbrace{G_{2}\times G_2\times\cdots\times G_2}_{r}$ be abelian group where $G_2=<a:a^2=1>$. Set $s=2r$, we assume that $s\mid(q-1)$. Let $\widehat{G}$ be the set of characters for $G$ with respect to  $\mathbb{F}_{q}$. Then $\widehat{G}=\{\chi_0,\chi_1,\ldots,\chi_{s-1}\}$ where $\chi_0,\chi_1,\ldots,\chi_{s-1}$ are the irreducible characters of $G$.
Therefore, the character table of $G$ is
$$ A=\begin{pmatrix}\chi_{0}(g_0)&\chi_{1}(g_0)&\cdots&\chi_{s-1}(g_0)\\\chi_{0}(g_1)&\chi_{1}(g_1)&\cdots&\chi_{s-1}(g_1)\\ \vdots &\vdots&\cdots& \vdots\\\chi_{0}(g_{s-1})&\chi_{1}(g_{s-1})&\cdots&\chi_{s-1}(g_{s-1})\end{pmatrix}.$$
where $g_0,g_1,\ldots,g_{s-1}\in G$. From above to see that  $(A^{-1})^{T}=\frac{1}{s}A$.
\end{rem}
Now we would like to use matrix-product codes to construct LCD codes.

\begin{cor}  Let $l^{2}\equiv 1~\mathrm{mod}~4$  where $l$ is an odd prime power, and let $C_1 ,C_2,C_3,C_4$  be Hermitain LCD  codes over $\mathbb{F}_{l^{2}}$ with
parameters $[n,k_i ,d_i ]_{l^{2}}$. Then there exists a Hermitian LCD code over $\mathbb{F}_{l^{2}}$ with parameters $[4n,k_1 + k_2+k_3+k_4,\geq min \{4d_{1},2d_{2},2d_{3},d_{_4}\}]_{l^{2}}$.
\end{cor}
\pf  Suppose that $G=G_{2}\times G_{2}=\{(1,1),(x,1)(1,y),(x,y):x^2=y^2=1\}$. Then the character table of $G$ is
$$ A=\begin{pmatrix}1&1&1&1\\1&1&-1&-1\\ 1&-1&1&-1\\1&-1&-1&1\end{pmatrix}.$$
By above construction of the matrix $A$, Theorem 3.3, and Lemma 2.3, we know that the matrix product code $C=[C_{1},C_{2},C_{3},C_{4}]A$ is Hermitain LCD with parameters $[4n,k_1 + k_2+k_3+k_4,\geq min \{4d_{1},2d_{2},2d_{3},d_{_4}\}]_{l^{2}}$.
\qed

Similarly, We have the following Corollary.
\begin{cor}  Let $q\equiv 1~\mathrm{mod}~4$, and let  $C_1 ,C_2,C_3,C_4$  be  LCD  codes over $\mathbb{F}_{q}$ with
parameters $[n,k_i ,d_i ]_{q}$. Then there exists a MPLCD code over $\mathbb{F}_{q}$ with parameters $[4n,k_1 + k_2+k_3+k_4,\geq min \{4d_{1},2d_{2},2d_{3},d_{_4}\}]_{q}$.
\end{cor}

\begin{exam}
Let $p=5$. Then
$$ x^{12}-1=(x+1)(x+2)(x+3)(x+4)(x^{2}+x+1)(x^{2}+2x+4)(x^2+3x+4)(x^2+4x+1),$$
is the factorization of $x^{12}-1$ into irreducible factors over $\mathbb{F}_{5}$. It follows from Proposition 2.6 that $C_1=C_2=C_3=\langle x+1\rangle $ and $C_4=\langle (x+1)(x^2+2x+4)(x^2+3x+4)\rangle$ are all LCD codes over $\mathbb{F}_{5}$ with parameters $[12,11,2]_5$ and $[12,7,4]_5$ respectively. By Corollary 4.3, matrix-product code $C=[C_1,C_2,C_3,C_4]\begin{pmatrix}1&1&1&1\\1&1&-1&-1\\ 1&-1&1&-1\\1&-1&-1&1\end{pmatrix}$ is an LCD code  over $\mathbb{F}_{5}$ with parameters $[48,40,\geq 4]_5$ .
\end{exam}
\begin{exam}
Let $p=5$. Then
$$ x^{13}-1=(x+4)(x^4+x^3+4x^{2}+x+1)(x^4+2x^3+x^{2}+2x+1)(x^4+3x^3+3x+1),$$
is the factorization of $x^{13}-1$ into irreducible factors over $\mathbb{F}_{5}$. It follows from Proposition 2.6 that $C_1=\langle x+1\rangle $, $C_2=C_3=\langle x^4+x^3+4x^2+x+1\rangle$ and $C_4=\langle (x+4)(x^4+x^3+4x^{2}+x+1)(x^4+2x^3+x^{2}+2x+1)\rangle$ are all LCD codes over $\mathbb{F}_{5}$ with parameters $[13,12,2]_5$, $[13,9,4]_5$ and $[13,4,8]_5$ respectively. By Corollary 4.3, matrix-product code $C=[C_1,C_2,C_3,C_4]\begin{pmatrix}1&1&1&1\\1&1&-1&-1\\ 1&-1&1&-1\\1&-1&-1&1\end{pmatrix}$ is an LCD code over $\mathbb{F}_{5}$ with parameters $[52,34,\geq 8]_5$.
\end{exam}
\begin{exam}
Let $p=5$. Then
$$ x^{18}-1=(x+1)(x+4)(x^2+x+1)(x^2+4x+1)(x^6+x^3+1)(x^6+4x^3+1),$$
is the factorization of $x^{18}-1$ into irreducible factors over $\mathbb{F}_{5}$. It follows from Proposition 2.6 that $C_1=\langle x+1\rangle $, $C_2=C_3=\langle (x+1)(x^6+x^3+1)\rangle$ and $C_4=\langle (x^2+x+1)(x^2+4x+1)(x^6+x^3+1)(x^6+4x^3+1)\rangle$ are all LCD codes over $\mathbb{F}_{5}$ with parameters $[18,17,2]_5$, $[18,11,4]_5$ and $[18,2,9]_5$ respectively. By Corollary 4.3, matrix-product code $C=[C_1,C_2,C_3,C_4]\begin{pmatrix}1&1&1&1\\1&1&-1&-1\\ 1&-1&1&-1\\1&-1&-1&1\end{pmatrix}$ is an LCD code  over $\mathbb{F}_{5}$ with parameters $[72,41,\geq 8]_5$.
\end{exam}

\begin{exam}
Let $p=13$. Then
$$ x^{10}-1=(x+1)(x+12)(x^4+12x^3+x^{2}+12x+1)(x^4+x^3+x^{2}+x+1),$$
is the factorization of $x^{10}-1$ into irreducible factors over $\mathbb{F}_{13}$. It follows from Proposition 2.6 that $C_1=\langle (x^4+x^3+x^{2}+x+1)\rangle $, $C_2=C_3=\langle (x^4+12x^3+x^{2}+12x+1)(x^4+x^3+x^{2}+x+1)\rangle$ and $C_4=\langle (x+1)(x^4+12x^3+x^{2}+12x+1)(x^4+x^3+x^{2}+x+1)\rangle$ are all LCD codes over $\mathbb{F}_{13}$ with parameters $[10,6,2]_{13}$, $[10,2,5]_{13}$ and $[10,1,10]_{13}$ respectively. By Corollary 4.3, matrix-product code $C=[C_1,C_2,C_3,C_4]\begin{pmatrix}1&1&1&1\\1&1&-1&-1\\ 1&-1&1&-1\\1&-1&-1&1\end{pmatrix}$ is an LCD code  over $\mathbb{F}_{13}$ with parameters $[40,11,\geq 8]_{13}$, and by Theorem 3,2, matrix-product code $C=[C_1,C_2,C_2,C_4]\begin{pmatrix}1&0&0&0\\1&1&0&0\\ 1&1&1&0\\1&1&1&1\end{pmatrix}$ is an LCD code over $\mathbb{F}_{13}$  with parameters $[40,11,\geq 2]_{13}$ .
\end{exam}
\begin{exam}
Let $p=5^2$, $\omega$ is a 25-th primitive root of unity. Then
$$ x^{17}-1=(x+4)(x^8+w^4x^7+w^{21}x^6+w^{7}x^5+w^{13}x^4+w^7x^3+w^{21}x^{2}+w^4x+1)$$$$~~~~~~(x^8+w^{20}x^7+w^9x^6+w^{11}x^5+w^{17}x^4+w^{11}x^3+w^{9}x^{2}+w^{20}+1).$$
is the factorization of $x^{17}-1$ into irreducible factors over $\mathbb{F}_{5^2}$.  It follows from Proposition 2.9 that $C_1=\langle x^8+w^4x^7+w^{21}x^6+w^{7}x^5+w^{13}x^4+w^7x^3+w^{21}x^{2}+w^4x+1\rangle $, $C_2=C_3=\langle (x+4)(x^8+w^4x^7+w^{21}x^6+w^{7}x^5+w^{13}x^4+w^7x^3+w^{21}x^{2}+w^4x+1)\rangle$ and $C_4=\langle (x^8+w^4x^7+w^{21}x^6+w^{7}x^5+w^{13}x^4+w^7x^3+w^{21}x^{2}+w^4x+1)(x^8+w^{20}x^7+w^9x^6+w^{11}x^5+w^{17}x^4+w^{11}x^3+w^{9}x^{2}+w^{20}+1)\rangle$ are all Hermitain LCD codes over $\mathbb{F}_{5^2}$ with parameters $[17,9,9]_{5^2}$, $[17,8,10]_{5^2}$ and $[17,1,17]_{5^2}$ respectively. By Corollary 4.3, matrix-product code $C=[C_1,C_2,C_3,C_4]\begin{pmatrix}1&1&1&1\\1&1&-1&-1\\ 1&-1&1&-1\\1&-1&-1&1\end{pmatrix}$ is a Hermitain LCD code over $\mathbb{F}_{5^2}$ with parameters $[68,26,\geq17]_{5^2}$, and by Theorem 3,4, matrix-product code $C=[C_1,C_1,C_2,C_2]\begin{pmatrix}1&0&0&0\\1&1&0&0\\ 1&1&1&0\\1&1&1&1\end{pmatrix}$ is a Hermitain LCD code  over $\mathbb{F}_{5^2}$ with parameters $[68,34,\geq 9]_{5^2}$ .
\end{exam}

\begin{thm} Let $C_1$ and $C_2$ be LCD codes over $\mathbb{F}_2$ with parameters $[n, k_1, d_1 ]_2$ and
$[n,k_2, d_2]_2$ respectively. If $A=\begin{pmatrix}1&0&1\\1&1&0\\ 1& 1&1\end{pmatrix}$,  then matrix-product code
$C=[C_1 ,C_1 ,C_2]A$ is MPLCD over $\mathbb{F}_{2}$ with  parameters $[3n,2k_1 + k_2,\geq min \{2d_{1},d_{2}\}]_{2}$.
\end{thm}
\pf Since $A=\begin{pmatrix}1&0&1\\1&1&0\\ 1& 1&1\end{pmatrix}$, we have $(A^{-1})^{T}=\begin{pmatrix}1&1&0\\1&0&1\\ 1& 1&1\end{pmatrix}$.
Hence,
$$C^{\perp}=[C_1^{\perp},C_1^{\perp},C_2^{\perp}]\begin{pmatrix}1&1&0\\1&0&1\\ 1& 1&1\end{pmatrix}.$$

Suppose that $\eta \in C\cap C^{\perp}$. Then there exist $c_1,c_1^{'}\in C_{1},c_2\in C_{2}$ and $c_1^{\perp},(c_1^{'})^{\perp}\in C_{1}^{\perp},c_2^{\perp}\in C_{2}^{\perp}$ such that $(c_1+c_1^{'}+c_2,c_1^{'}+c_2,c_1+c_2)=(c_1^{\perp}+(c_1^{'})^{\perp}+c_2^{\perp},c_1^{\perp}+c_2^{\perp},(c_1^{'})^{\perp}+c_2^{\perp})$ by the definition of matrix-product code. This implies that $$c_1+c_1^{'}+c_2=c_1^{\perp}+(c_1^{'})^{\perp}+c_2^{\perp},$$
$$c_1^{'}+c_2=c_1^{\perp}+c_2^{\perp},~~~~~~~~~~~~~~~~$$
$$c_1+c_2=(c_1^{'})^{\perp}+c_2^{\perp}.~~~~~~~~~~~~$$

By $c_1^{'}+c_2=c_1^{\perp}+c_2^{\perp}, c_1+c_2=(c_1^{'})^{\perp}+c_2^{\perp}$, and $C_1,C_1^{\perp}$ linearly, we have $c_1-c_1^{'}=(c_1^{'})^{\perp}+c_1^{\perp}\in C_1\cap C_1^{\perp}=\{0\}$, i.e., $c_1=c_1^{'},(c_1^{'})^{\perp}=c_1^{\perp}$. On other hand, by $c_1+c_1^{'}+c_2=c_1^{\perp}+(c_1^{'})^{\perp}+c_2^{\perp}$, we obtain $c_2=c_2^{\perp}\in C_2\cap C_2^{\perp}=\{0\}$, i.e., $c_2=c_2^{\perp}=0$, which prove $c_1=c_1^{'}=0$. Therefore, $\eta=0$. This means that $C$ is a matrix-product LCD code over $\mathbb{F}_{2}$.
\qed
\begin{exam}
Let $p=2$. Then
$$ x^{10}-1=(x+1)^2(x^4+x^3+x^{2}+x+1)^2.$$
is the factorization of $x^{10}-1$ into irreducible factors over $\mathbb{F}_{2}$. It follows from Proposition 2.7 that $C_1=\langle (x+1)^2\rangle $, $C_2=\langle (x^4+x^3+x^{2}+x+1)^2\rangle$ are LCD codes over $\mathbb{F}_{2}$ with parameters $[10,8,2]_2$ and $[10,2,5]_{2}$  respectively. By Theorem 4.4, matrix-product code $C=[C_1,C_1,C_2]\begin{pmatrix}1&0&1\\1&1&0\\ 1& 1&1\end{pmatrix}$ is an LCD code over $\mathbb{F}_{2}$ with parameters $[30,18,\geq 4]_{2}$ .
\end{exam}

\section{Conclusion}
We have developed  new methods of constructing LCD codes from matrix-product codes over finite field $\mathbb{F}_{q}$. Using those methods, we have constructed  good LCD
codes. We believe that matrix-product codes over finite field $\mathbb{F}_{q}$ will be a good source for constructing good LCD codes. In a future work, we will use the computer algebra system MAGMA to find more good LCD  codes.

\textbf{Acknowledgements}

This work was supported by Research Funds of Hubei Province, Grant
No. D20144401.

\end{document}